\documentclass[a4paper,10pt]{article}
\usepackage{amsmath,amsfonts}
\usepackage{times}
\usepackage[T1]{fontenc}
\usepackage{epsfig}
\usepackage{amssymb}

\begin{document}

\title{\Huge{\bf Investigation of restricted baby Skyrme models}}

\author{C. Adam$^{a)}$\thanks{adam@fpaxp1.usc.es}, T. Roma\'{n}czukiewicz 
$^{b)}$\thanks{romanczukiewicz@th.if.uj.edu.pl},
J. S\'{a}nchez-Guill\'{e}n
$^{a)c)}$\thanks{joaquin@fpaxp1.usc.es} \\
A. Wereszczy\'{n}ski
$^{b)}$\thanks{wereszczynski@th.if.uj.edu.pl}
       \\
       \\ $^{a)}$ Departamento de F\'isica de Part\'iculas, Universidad
       \\ de Santiago, and Instituto Galego de F\'isica de Altas Enerx\'ias
       \\ (IGFAE) E-15782 Santiago de Compostela, Spain
       \\
       \\ $^{b)}$ Institute of Physics,  Jagiellonian University,
       \\ Reymonta 4, Krak\'{o}w, Poland
       \\
       \\ $^{c)}$ Sabbatical leave at: Departamento de F\'isica Te\'orica,
       \\ Universidad de Zaragoza, 50009 Zaragoza, Spain}

\maketitle

\begin{abstract}
A restriction of the baby Skyrme model consisting of the quartic and 
potential terms only is investigated in detail for a wide range of potentials. 
Further, its properties are compared with those of the 
corresponding full baby Skyrme models. 
We find that topological (charge) as well as 
geometrical
(nucleus/shell shape) features of baby skyr\-mions are captured already by
the soliton solutions of the restricted model. Further, we find a
coincidence between the compact or non-compact nature of solitons in the 
restricted model, on the one hand, and
the existence or non-existence of multi-skyrmions in the full baby Skyrme 
model, on the other hand.

\end{abstract}
\newpage
%%%%%%%%%%%%%%%%%%%%%%%%%%%%%%%%%%%%%%%%%%%%%%%%%%%%%%%%%%%%%%
\section{Introduction}
%%%%%%%%%%%%%%%%%%%%%%%%%%%%%%%%%%%%%%%%%%%%%%%%%%%%%%%%%%%%%
The baby Skyrme model was introduced originally as a planar
analogue of the three-dimensional Skyrme model \cite{skyrme}. Also
its target space is simplified accordingly ($S^2$ instead of the
SU(2) target space of the Skyrme model) such that static field
configurations can be classified according to their winding number
in both cases. Like the Skyrme model, also the baby Skyrme model
consists of a quadratic kinetic term (the O(3) nonlinear sigma
model term), a quartic kinetic term (the analogue of the Skyrme
term), and a potential. For the baby Skyrme model, the inclusion
of a potential term is obligatory for the existence of static
finite energy solutions. The  specific form of this potential term
is, however, quite arbitrary, and different potentials have been
studied \cite{old} - \cite{other} (for some recent studies see
e.g. \cite{karliner}, \cite{comp-bS}). In addition to its role as
a toy model for the Skyrme model, the baby Skyrme model has also
found some independent applications in condensed matter physics in
the description of the quantum Hall effect \cite{qhe}.

The energy functional of the baby Skyrme model for static
configurations is
\begin{equation} \label{bS}
E = \frac{1}{2} \int d^2 x \left( \partial_i \vec \phi \cdot
\partial_i \vec \phi + \frac{1}{4} (\epsilon_{ij}\partial_i \vec
\phi \times \partial_j \vec \phi )^2 + \mu^2 V(\vec \phi) \right)
\end{equation}
where $\vec \phi $ is a three-component vector field with unit
modulus $|\vec \phi |=1$, and $V$ is the potential. Further, $\mu$
is a positive constant.

Although the baby Skyrme model, due to its simpler structure, may offer a 
way for
a better understanding of the solutions in the $(3+1)$ Skyrme model,
it is still a non-integrable, highly complicated, topologically non-trivial,
non-linear field theory. Many properties of baby skyrmions are established 
mainly
by numerical simulations, whereas it is much more difficult to gain 
analytical understanding. Therefore,
it is important to know whether there exists the possibility for any further
simplification which would keep us in the class of Skyrme-like models and, 
nevertheless, allows
for some exact analytical calculations. For example, one may try to identify 
which
features of the solutions of the baby Skyrme model are governed by which 
part of the model.
Then, neglecting a particular part of the Lagrangian,
one could investigate a simplified model. In case of static solutions only two 
simplifications are possible. One may either suppress both the potential and 
the quartic, pure Skyrme term, which leads to the scale invariant O(3) sigma
model with its well-known meromorphic solutions. Or one may opt for an energy
functional which is not scale invariant. 
In this case, both the
quartic kinetic term and the potential are mandatory for the
evasion of Derrick's theorem, whereas the quadratic term is optional from
this point of view. The latter fact led Gisiger and Paranjape to
consider the model with the quadratic term omitted \cite{GP}, and
with the so-called ``old'' potential
\begin{equation}
V(\vec \phi) = (\hat n - \vec \phi)^2 = 2(1-\hat n \cdot \vec
\phi) \label{old}
\end{equation}
where $\hat n$ is a constant unit vector which selects the vacuum.
The resulting model has a huge amount of symmetry. Indeed, both
the area-preserving diffeomorphisms on base space \cite{GP} and an
abelian subgroup of the area-preserving difeomorphisms on target
space \cite{ab-dif} are symmetries of the static GP (=Gisiger and
Paranjape) model (as the model consists of a potential term and the pure
Skyrme term only, we will also call it the ``pure baby Skyrme model'').
Further, already for a spherically symmetric
ansatz, GP found an infinite number of exact soliton solutions,
which, together with the huge symmetry,  points towards
the integrability of this model in $(2+0)$ dimensions \cite{integr}. These 
solitons are, in fact,
compact, that is, they differ from the vacuum only in a finite
compact region of the (two-dimensional) base space \cite{rosenau}. A further
explanation of their compact nature can be found in
\cite{comp-bS}, where it is demonstrated that the GP model can, in
fact, be mapped to the signum-Gordon model (the latter model is
known to support compact solutions, see \cite{comp}). In addition,
GP derived a Bogomolny bound in their model, but the solutions
they found do not saturate this bound.

The main aim of the present work is to understand correlations between
properties of soliton solutions in the pure and the full
baby Skyrme models. It occurs that some important topological as well as
geometrical features of baby skyrmions can be captured already
by solutions of the simplified model.

First, we show that the previously proven integrability
of the static version of the pure baby Skyrme models \cite{integr} can be 
extended to the $(2+1)$
space-time case. However, now the models are integrable in the sense of 
generalized integrability \cite{alvarez}. In particular,
we construct an infinite set of conserved quantities.

In a next step, we generalize the results of GP by studying pure baby
Skyrme models with different potentials. We find that, depending on the
type of potential we choose, there exist soliton solutions with
qualitatively different behavior. In some cases, there exist
compact solitons, whereas in other cases the solitons have an
exponential tail or are even only localized by some inverse powers
of the radius. Further, we find an interesting coincidence between
the behavior of the solitons in the GP like model (with only a
quartic kinetic term), on the one hand, and the existence of
multi-soliton solutions of the full baby Skyrme model, on the
other hand. Multi-solitons in the full model seem to exist only
for potentials such that the GP model has compacton solutions,
whereas they do not exist if the solitons of the corresponding GP
model have an exponential or power-like tail. Moreover, also
the nucleus respectively shell-like structure is preserved when
we go from the pure to the full baby Skyrme model.

We also comment on the issue of Bogomolny bounds.
Specifically, we point out that there exists a second Bogomolny
bound in the GP model, and all exact soliton solutions saturate
this bound. Further, this bound continues to hold for GP like
models with arbitrary potential. This bound was originally found
in \cite{deI-W}, as a contribution to an improved Bogomolny bound
for the full baby Skyrme model. We just add the information that
in GP like models this bound is, in fact, saturated, and solutions
are, therefore, solutions of the corresponding Bogomolny
equations.
Further, this second Bogomolny bound of the
GP model immediately leads to a second, improved Bogomolny bound for the
full baby Skyrme model \cite{deI-W}. Solutions of the full baby
Skyrme model do not saturate this bound for general potentials.
This second bound is, however, much tighter than the more widely
known Bogomolny bound which stems from the quadratic O(3) sigma
model term alone. In addition, the new bound becomes very tight in
the limits of very weak or very strong potential. We explore the
quality (i.e. tightness) of the bound for several parameter
choices. We remark that an equivalent Bogomolny bound may be derived for
generalized Skyrme models in three dimensions \cite{baryons}. There the
pertinent topological charge is the baryon number, and the limiting model
which saturates the Bogomolny bound gives an accurate description of basic
properties of nuclei, like e.g. masses and sizes.

Finally, we comment on time-dependent, non-topological solutions in
the pure baby Skyrme model.
%%%%%%%%%%%%%%%%%%%%%%%%%%%%%%%%%%%%%%%%%%%%%%%%%%%%%%%%%%%%%%%%%%
\section{Generalized integrability in (2+1) dimensions}
%%%%%%%%%%%%%%%%%%%%%%%%%%%%%%%%%%%%%%%%%%%%%%%%%%%%%%%%%%%%%%%%%%
The model we are going to investigate in this work is defined by
the following Lagrange density
\begin{equation}
L= \frac{1}{2} \left( \frac{1}{2}
(\partial_{\mu} \vec \phi \times \partial_{\nu} \vec \phi )^2
+ \mu^2 \; V(\phi^3), \right) 
\end{equation}
where the potential $V$ is any potential depending only on the third component
of the unit vector field. 
In applications to the quantum Hall effect, this restricted model corresponds
to the limit of infinitely strong magnetic field.  
After using the standard stereographic
projection
$$
\vec{\phi }=\frac{1}{1+|u|^2} \left( u+\bar{u}, -i ( u-\bar{u}),
1 - |u|^2  \right)
$$
we derive equation of motion
$$
\partial_{\mu} \mathcal{K}^{\mu} - \frac{\mu^2}{4} \bar{u} (1+|u|^2)^2 V'=0,
$$
where ($u_\mu \equiv \partial_\mu u$, etc.)
$$
\mathcal{K}^{\mu}=\frac{K^{\mu}}{(1+|u|^2)^2}, \;\;\; 
K^{\mu}=(u_{\nu}\bar{u}^{\nu})\bar{u}^{\mu}
-\bar{u}_{\nu}^2u^{\mu}
$$
and the prime denotes differentiation with respect to $u\bar{u}$. The
construction of conserved quantities is performed by a well understood
procedure originally described in \cite{alvarez}. The conserved currents are
chosen as
\begin{equation}
J_{\mu}=\frac{\delta G}{\delta \bar{u}} \mathcal{K}_{\mu}-
\frac{\delta G}{\delta u} \bar{\mathcal{K}}_{\mu},
\;\;\;\; G=G(u\bar{u}),
\end{equation}
where $G$ is an arbitrary function which depends on the modulus of the complex 
field.
Obviously, we have infinitely many such currents as these functions are not 
restricted
to any particular form. Then, one may easily calculate the divergence 
of the currents and
get
$$
\partial_{\mu}J^{\mu}= \frac{\delta^2 G}{\delta \bar{u}^2} 
\mathcal{K}_{\mu}\bar{u}^{\mu}+
\frac{\delta^2 G}{\delta \bar{u} \delta u} \mathcal{K}_{\mu}u^{\mu}
+\frac{\delta G}{\delta \bar{u}} \partial^{\mu} \mathcal{K}_{\mu}
-\frac{\delta^2 G}{\delta u^2} \bar{\mathcal{K}}_{\mu}u^{\mu}-
\frac{\delta G}{\delta u \delta \bar{u}} \bar{\mathcal{K}}_{\mu} \bar{u}^{\mu}
-\frac{\delta G}{\delta u} \partial^{\mu} \bar{\mathcal{K}}_{\mu}.
$$
Using the properties of $\mathcal{K}_{\mu}$
$$\mathcal{K}_{\mu}\bar{u}^{\mu}=\bar{\mathcal{K}}_{\mu}u^{\mu}, \;\;\;  
\mathcal{K}_{\mu}u^{\mu}= \bar{\mathcal{K}}_{\mu} \bar{u}^{\mu}$$
we find
$$ \partial_{\mu}J^{\mu}=G' ( u\; \partial^{\mu} \mathcal{K}_{\mu} - 
\bar{u}\; \partial^{\mu} \bar{\mathcal{K}}_{\mu})=0,$$
where the last step follows from the equation of motion.
\\
In contrast to the pure baby Skyrme model, its full version is not 
integrable,
resulting in a rather complicated dynamics of the baby skyrmions. 
It has been shown, however,
that the full baby Skyrme model possesses an integrable submodel defined by
an additional constraint which must be imposed in addition to the 
equations of
motion, namely the eikonal equation 
$(\partial_{\mu} u)^2=0$. This constraint
is quite restrictive and it is not known wether this integrable 
submodel
can lead to interesting solutions. On the contrary, 
the pure baby Skyrme model is integrable 
(in the
specific meaning of the generalized integrability) without any additional
constraint.
\\
One may notice, that the pure baby Skyrme model explores an integrable
limit of the full baby model as was originally considered by 
Aratyn et.al. in the
case of the Skyrme-Faddeev-Niemi model \cite{aratyn2}. Indeed, their limit
$m^2 \rightarrow 0$ leads to a model without the quadratic part of the action.
However, as they omitted the potential term, as well, the limit gave a
theory without static soliton solutions, which probably
was the reason why they
did not investigate that model in detail.
%%%%%%%%%%%%%%%%%%%%%%%%%%%%%%%%%%%%%%%%%%%%%%%%%%%%%%%%%%%%%%%%%%
\section{Solitons in the pure baby Skyrme model}
%%%%%%%%%%%%%%%%%%%%%%%%%%%%%%%%%%%%%%%%%%%%%%%%%%%%%%%%%%%%%%%%%%
%%%%%%%%%%%%%%%%%%%%%%%%%%%%%%%%%%%%%%%%%%%%%%%%%%%%%%%%%%%%%
\subsection{One-vacuum potentials}
%%%%%%%%%%%%%%%%%%%%%%%%%%%%%%%%%%%%%%%%%%%%%%%%%%%%%%%%%%%%%
Let us consider the energy functional for the pure baby Skyrme
model
\begin{equation}
E = \frac{1}{2} \int d^2 x \left( \frac{1}{4}
(\epsilon_{ij}\partial_i \vec \phi \times \partial_j \vec \phi )^2
+ \mu^2 V(\phi^3) \right),
\end{equation}
where we specify the potential as
\begin{equation}
V= 4 \left( \frac{1-\phi^3}{2} \right)^s, \label{old s}
\end{equation}
which for $s=1$ gives the old baby Skyrme potential. Here $s >0$. Such a
family of generalized old baby potentials has been analyzed by
Karliner and Hen in the context of the rotational symmetry breaking in the
full baby Skyrme model \cite{karliner}.
\\
After the stereographic projection we get
$$
E = \int d^2 x \left( 2\frac{(\nabla u \nabla \bar{u})^2-(\nabla
u)^2(\nabla \bar{u})^2}{(1+|u|^2)^4}+ 2 \mu^2\left(
\frac{|u|^2}{1+|u|^2} \right)^s \right),
$$
Further, we assume the symmetric ansatz
\begin{equation}
u=e^{i  n\varphi} f(r), \label{ansatz}
\end{equation}
which is compatible with the equations of motion, and introduce a
new function
$$
1-g=\frac{1}{1+f^2},
$$
together with a new variable $y=r^2/2$. For topologically
non-trivial solution one has to impose the following boundary
conditions
$$
f(r=0)=\infty, \;\;\; f(r=R)=0, \;\;\; f'(r=R)=0
$$
where $R$ is finite for compactons or $R=\infty$ for standard
solitons (then, the third expression does not lead to a new
condition). In terms of the new function $g$ we get
$$
g(y=0)=1, \;\;\; g(y=Y)=0, \;\;\; g'(y=Y)=0,
$$
Now, the energy takes the simply form
$$
E = 2\pi \int d y \left( 2n^2 g_y^2+ 2 \mu^2 g^s \right).
$$
The finite energy solutions must obey the pertinent first order
Bogomolny equation which is easily obtained from the second order equation
of motion,
$$
n^2 g_y^2 = \mu^2 g^s.
$$
This simple equation supports topological solutions (i.e.,
solutions with the previously specified boundary conditions), which
may be divided into three classes.
\\
{\it (i)} $s \in (0,2)$ {\it - Compactons}
\\
The pertinent solutions are compactons
\begin{equation}
g(y)=\left\{
\begin{array}{cc}
\left( 1 - \frac{y}{y_0} \right)^{\frac{2}{2-s}} & 0 \leq y \leq
y_0
\\
 &
\\
0 & y \geq y_0,
\end{array}
\right.
\end{equation}
where the boundary of the compacton is located at
$$
y_0=\frac{2n}{\mu(2-s)}.
$$
The approach to the vacuum becomes more and more rapid as we tend to
$s=2$. Such compact solutions represent rotationally symmetric
multi-solitons configuration which can be understood as a
collection of $n$ $Q=1$ solitons located on top of each other.
However, due to the compacton nature of the solutions one may
easily construct a non-rotational solution with $Q=n$ by a
collection of some solutions with lower topological charges,
provided that they are sufficiently separated and the sum of
charges is equal to $n$.
\\
{\it (ii)} $s=2$ {\it - Exponentially localized solitons}
\\
Now, we get a soliton with the standard exponential tails
\begin{equation}
g(y)=e^{-\frac{\mu y}{n}}.
\end{equation}
Again, higher charge solitons may be interpreted as several $Q=1$ solitons
sitting on top of each other. Further, it is no longer obvious how to
construct non-symmetric solutions and solutions corresponding to
well-separated solitons. We remark that this case has already been
investigated in \cite{Tchr 3}.
\\
{\it (iii)} $s>2$ {\it - Power-like localized solitons}
\\
The third class of solutions is formed by solitons with
power-like tails
\begin{equation}
g(y)= \left( \frac{y_0}{y_0+y}\right)^{\frac{2}{s-2}},
\end{equation}
where
$$
y_0=\frac{2n}{\mu(s-2)}.
$$
Three remarks are appropriate.
\\
Firstly, as one could expect, in the case of solutions of a first order
Bogomolny equation, for all models the total energy is
proportional to the topological charge
$$E= \frac{16\pi}{s+2} \mu |Q|.$$
Secondly, from a geometrical point of view all these solutions share a
common feature. Namely, they are of nucleus type with their energy density
concentrated around the origin. This resembles the baby skyrmions in the 
old baby
model, where solitons also possess such a nucleus shape.
\\
Finally, there is a very intriguing coincidence between the compactness
of soliton solutions of the pure model and the existence, for some values of
the parameters, of multi baby-skyrmions in the full model. In fact, for
potentials with $s \in (0,2)$ solitons in the pure model are of the compact 
type.
On the other hand, as is known from previous numerical works, full
baby Skyrme models with the same potentials support multi
soliton solutions. Moreover, for baby Skyrme models with potentials with 
$s \geq 2$, no
multi-soliton configurations have been found. 
It corresponds, on the pure model side,
to non-compact topological solutions 
(exponentially or power-like localized).
It seems as if the instability of the multi-skyrmions in the full baby models 
has its origin
in the non-compactness of the solitons in the pure models. Equivalently, one
may conjecture that compactons in the pure models are seeds for
multi-skyrmions in the full theory.
%%%%%%%%%%%%%%%%%%%%%%%%%%%%%%%%%%%%%%%%%%%%%%%%%%%%%%%%
\subsection{Two-vacua potentials}
%%%%%%%%%%%%%%%%%%%%%%%%%%%%%%%%%%%%%%%%%%%%%%%%%%%%%%%%
%%%%%%%%%%%%%%%%%%%%%%%%%%%%%%%%%%%%%%%%%%%%%%
\subsubsection{Generalized new baby potential}
%%%%%%%%%%%%%%%%%%%%%%%%%%%%%%%%%%%%%%%%%%%%%%
Another type of potentials we are going to analyze is given by the
expression
\begin{equation}
V= 4 \left( \frac{1-(\phi^3)^2}{4} \right)^s,
\end{equation}
with $s >0$. In the limit $s=1$ one can recognize the new baby
potential, a well-known potential with two vacua $\phi^3=\pm 1$. The
pertinent first order equation of motion reads
$$
n^2 g_y^2 = \mu^2 (g(1-g))^s.
$$
For this potential we find two classes of solutions.
\\
{\it (i)} $s \in (0,2)$ {\it - Compactons}
\\
The general solution describes a compact ring. In fact, both vacua
are of the compacton type i.e., a solution tends to them in a
power like manner. More precisely, the solution (not in explicit
form) reads
\begin{equation}
\frac{2}{2-s} _2F_1[1-\frac{s}{2},
\frac{s}{2},2-\frac{s}{2},g]=\frac{\mu(y+y_0)}{n},
\end{equation}
where $_2F_1$ is a hypergeometric function. An example of an
explicit solution can be found for $s=1$
\begin{equation}
g(y)=\left\{
\begin{array}{cc}
\left(\sin \frac{2\mu (y-y_0)}{n} \right)^2 & 0 \leq y \leq y_0 \\
 &
\\
0 & y \geq y_0
\end{array}
\right.
\end{equation}
and
$$
y_0=\frac{n \pi}{4 \mu }.
$$
The corresponding energy density is not concentrated at the origin
but has its maximum at $y_{max}=n\pi/8\mu $, forming a ring-like
structure. This solution can be easily extended to a family of
 ring-like solutions with the maximum of the energy density located at any
$y \geq y_{max}$. Again, all solutions lead to a linear
energy-charge relation.
\\
Let us underline that the fact that solutions are of the ring type again 
resembles the
geometric properties of soliton solutions in the full baby Skyrme model.
Indeed, the baby skyrmions in the new baby Skyrme model have this property.
\\
{\it (ii)}  $s \geq 2$ {\it - No solitons}
\\
It is easy to check that there are no topological solitons for
such a value of the parameter. The reason for this is that both
vacua are approached exponentially $(s=2)$ or polynomially
($s>2$). It is of course impossible to join both vacua in such a
way on the semi-infinite segment $y\in [0,\infty)$.
%%%%%%%%%%%%%%%%%%%%%%%%%%%%%%%%%%%%%%%%%%%%%%%%%%
\subsubsection{Generalized new baby potential - non-symmetric case}
%%%%%%%%%%%%%%%%%%%%%%%%%%%%%%%%%%%%%%%%%%%%%%%%%%
\noindent The last family of potentials of this type we want to
analyze has the form
\begin{equation}
V= 4 \left( \frac{1+\phi^3}{2} \right)^a \left( \frac{1-\phi^3}{2}
\right)^b,
\end{equation}
with $a,b > 0$. The properties of the solutions are as follows.
\\
{\it (i) $a \in (0,2)$ and $b \in (0,2)$} {\it - Compactons}
\\
We get compactons as the approach to both vacua is in a power-like
manner.
\\
{\it (ii) $a =2$ and $b \in (0,2)$ or $a \in (0,2)$ and $b =2$}
{\it - Exponentially localized solitons}
\\
We get standard soliton
solutions with exponential tails. As an example, we get
\begin{equation}
g(y)= \frac{2}{1+\cosh \frac{\mu y}{n}}
\end{equation}
for the first case ($a=2$, $b=1$), and
\begin{equation}
g(y)= \left( \tanh \frac{\mu y}{2 n } \right)^2,
\end{equation}
for  second case ($a=1$, $b=2$), respectively. Notice that the
second solution possesses the reversed boundary conditions, i.e.,
$g(0)=0$, $g(\infty)=1$.
\\
{\it (iii) $a >2$ and $b \in (0,2)$ or $a \in (0,2)$ and $b > 2$}
{\it - Power-like localized solitons}
\\
We get solitons with power-like tails. As an example one may
consider the choice of parameters $a=4, b=1$. Then, the pertinent
solution reads
\begin{equation}
\sqrt{\frac{1}{g^2} - \frac{1}{g}} + \mbox{ar cosh}
\frac{1}{\sqrt{g}}=y.
\end{equation}
All presented solutions can be extended to shell-like
configurations. As before one gets that the energy is proportional
to the topological charge.
\\
{\it (iii) $a \geq 2$ and $b\geq 2$} {\it - No solitons}
%%%%%%%%%%%%%%%%%%%%%%%%%%%%%%%%%%%%%%%%%%%%%%%%%%%%%%%
\subsection{Vortex potentials}
%%%%%%%%%%%%%%%%%%%%%%%%%%%%%%%%%%%%%%%%%%%%%%%%%%%%%%%
A different family of potentials also considered for the baby
Skyrme model is given by the following expression
\begin{equation}
V= 4 \left( (\phi^3)^2 \right)^s \left( \frac{1-\phi^3}{2} \right),
\end{equation}
where $s >0$. 
Here the vacuum manifold is completely different from the cases considered
up to now. 
It still consists of two disconnected components, but now only one of
the two components is a point (the north pole $\phi^3 =1$ of the target space
$S^2$), whereas the other component of the vacuum is a circle $S^1$: $\phi^3
=0$, $(\phi^1)^2 + (\phi^2)^2 =1$. In our parametrization, this vacuum
corresponds to $g=1/2$. 
The presence of this new vacuum allows for field configurations which approach
it in the limit $|\vec x|\to \infty $. Such fields 
may be divided into
disjoint topological classes as
$$\vec{n}_{\infty}: \mathbb{R}^2_{r\rightarrow \infty} \simeq  \mathbb{S}^1 
\longrightarrow U(1)\simeq \mathbb{S}^1.$$
The relevant topological index is the winding number $Q \in
\pi_1(S^1)$, and the corresponding field configurations are
vortices. The important question is, of course, whether such vortex
configurations are possible solutions of the underlying field theory. In this
respect, the full and the pure baby Skyrme model are fundamentally different.
The reason is that the quadratic O(3) sigma model term, which is present only
in the full model, produces 
infinite energy for vortex configurations. Indeed, in
polar coordinates the static O(3) term essentially is the sum of a radial
gradient squared and an angular gradient squared. In the limit, the angular
gradient behaves like $r^{-1}\hat e_\varphi \partial_\varphi \exp (in\varphi)
=in r^{-1}\hat e_\varphi$ where $n$ is the vortex number, so for nonzero $n$
the energy density goes like $r^{-2}$ and produces a logarithmic divergency.
As a consequence, there are no vortices with finite energy in the full baby
Skyrme model. In the pure model, on the other hand, only the quartic term is
present in addition to the potential. Further, due to its antisymmetry, {\em
  both} gradients have to appear in this term (in fact, both 
quadratically). Therefore, the slowly decaying angular gradient is always
multiplied by a rapidly decaying radial gradient, such that the resulting
energy density decays sufficiently fast to have finite energy. In short,
finite energy vortices are not excluded in the pure model, and they turn out
to exist, as we shall see in a moment.
\\
{\it (i)} $s \in (0,1)$ {\it - Vortex compactons}
\\
We get compact vortices. As an example with explicit solutions let
us consider the model with $s=1/2$. Then,
\begin{equation}
g(y)=\left\{
\begin{array}{cc}
\frac{1}{2} \left( \cosh \frac{4\sqrt{2}(y-y_0)}{n} \right)^2 & y
\leq y_0 \\
&
\\
\frac{1}{2} & y > y_0,
\end{array} \right.
\end{equation}
or
\begin{equation}
g(y)=\left\{
\begin{array}{cc}
 \frac{1}{2} \left( \sin
\frac{4\sqrt{2}y}{n} \right)^2 & y \leq \tilde{y}_0
\\
&
\\
\frac{1}{2} & y \geq \tilde{y}_0,
\end{array} \right.
\end{equation}
where $$ y_0=\frac{n}{4\sqrt{2}} \mbox{arc cosh} \sqrt{2}, \;\;\;
\tilde{y}_0=\frac{\pi n}{8\sqrt{2}}.$$ The first solution
represents vortex $(g \in (1,1/2))$ whereas the second anti-vortex
$(g \in (0,1/2))$. Quite interestingly, because of the compactness
of the solutions one may construct a configuration carrying a
non-trivial value of the topological charge by gluing in a proper way
together both profile functions (i.e., both vortices). Namely,
\begin{equation}
g(y)=\left\{
\begin{array}{cc}
\frac{1}{2} \left( \cosh \frac{4\sqrt{2}(y-y_<)}{n} \right)^2 & y
\in [0,y_<] \\
 &
\\
\frac{1}{2} & y \in [y_<, y_<+y_0]
\\
&
\\
\frac{1}{2} \left( \sin \frac{4\sqrt{2}(y-(y_>+y_0))}{n} \right)^2
& y
\in [y_<+y_0, y_>+y_0] \\
0 & y > y_>+y_0,
\end{array} \right.
\end{equation}
where $$ y_<=\frac{n}{4\sqrt{2}} \mbox{arc cosh} \sqrt{2}  $$  $$
y_>=\frac{\pi n}{8\sqrt{2}} (1+ \frac{2}{\pi} \mbox{arc cosh}
\sqrt{2}).
$$ Thus, two vortices with the same winding numbers $n$ form a
baby skyrmion with the topological charge $Q=n$. Therefore, for such
potentials a baby skyrmion seems to dissolve into a pair of $U(1)$
vortices, which now are the true constituent topological objects.
\\
{\it (ii)} $s \geq 1$ {\it - Exponential or polynomial vortices}
\\
A pertinent example is given by $s=1$ case. Then,
\begin{equation}
g(y) = \frac{1}{2} \left(\coth \frac{2(y+y_0)}{n} \right)^2
\end{equation}
or
\begin{equation}
g(y) = \frac{1}{2} \left(\tanh \frac{2y}{n} \right)^2,
\end{equation}
where $y_0$ obeys $$\coth \frac{2y_0}{n} = \sqrt{2}. $$ Of course, now these
vortex solutions cannot be glued together to
form a stable configuration with non-zero
topological charge.
%%%%%%%%%%%%%%%%%%%%%%%%%%%%%%%%%%%%%%%%%%%%%%%%%%%%%%%%%%%%%%%%%%%
\section{Bogomolny bounds}
%%%%%%%%%%%%%%%%%%%%%%%%%%%%%%%%%%%%%%%%%%%%%%%%%%%%%%%%%%%%%%%%%%%
In this section we prove that the solutions of the previous section
are indeed
of the Bogomolny type, without performing any dimensional
reduction. Moreover, we derive the pertinent Bogomolny bound and
show that our solutions saturate it. In fact, in an investigation
of the full baby Skyrme model this bound has been originally
reported in \cite{deI-W}.
Let us begin by reviewing briefly the original Bogomolny bound found by GP.
They studied the energy functional  
$$
E = \frac{1}{2} \int d^2 x \left( \frac{1}{4}
(\epsilon_{ij}\partial_i \vec \phi \times \partial_j \vec \phi )^2
+ \mu^2 (\hat n - \vec \phi )^2 \right) 
$$
where $\hat n$ is a constant unit vector selecting the vacuum.
Further, they derived the following Bogomolny bound
$$
E = \frac{1}{2} \int d^2 x \left( \left( \frac{1}{2} \epsilon_{ij}
\partial_i \vec
\phi \times \partial_j \vec \phi  \pm \mu (\hat n - \vec \phi )
\right)^2 \mp \mu (\hat n - \vec \phi )\cdot \epsilon_{ij}
\partial_i \vec \phi \times \partial_j \vec \phi \right)
$$
$$
=\frac{1}{2} \int d^2 x \left( \left( \frac{1}{2} \epsilon_{ij}
\partial_i \vec
\phi \times \partial_j \vec \phi  \pm \mu (\hat n - \vec \phi )
\right)^2 \pm \mu  \epsilon_{ij}  \vec \phi \cdot (\partial_i \vec
\phi \times \partial_j \vec \phi ) \right)
$$
where a total divergence term which gives zero upon integration
has been omitted in the last expression. The energy, therefore,
obeys a Bogomolny bound
$$
E \ge E_B \equiv 4\pi \mu |Q|
$$
were the integer-valued topological charge $Q$ is
$$
Q = \frac{1}{8\pi} \int d^2 x \epsilon_{ij}  \vec \phi \cdot
(\partial_i \vec \phi \times \partial_j \vec \phi ) .
$$
The solutions of GP (see also section 3.1), however, do not
saturate this Bogomolny bound. Instead, they satisfy
$$
E_{\rm sol} = \frac{4}{3} E_B
$$
so they are still proportional to the topological charge, just the
coefficient in front is slightly bigger than the coefficient for
the Bogomolny bound. Firstly, let us remark that nontrivial
solutions saturating this bound cannot exist. Indeed, the
corresponding Bogomolny equation is
$$
\frac{1}{2} \epsilon_{ij} \partial_i \vec \phi \times \partial_j
\vec \phi  \pm \mu (\hat n - \vec \phi ) =0
$$
and multiplying this by $\times \vec \phi$ one easily concludes
$$
\hat n \times \vec \phi =0
$$
and any solution $\vec \phi$ must be proportional to the trivial
constant vacuum vector $\hat n$.
\\
Secondly, the fact that the explicit solutions still are linear in
the topological charge, just with a slightly bigger prefactor,
already indicates that there might exist another, tighter
Bogomolny bound which is saturated by the energies of
the solutions.
We will demonstrate now that this is indeed the case (the bound itself has
already been found in \cite{deI-W}, as a contribution to an improved bound for
the full baby Skyrme model). For this
purpose, we observe first that the vectorial expression whose
square gives the quartic kinetic term necessarily is proportional
to the vector $\vec \phi$ itself and may therefore be written like
$$
\epsilon_{ij} \partial_i \vec \phi \times \partial_j \vec \phi =
\vec \phi \left( \vec \phi \cdot \epsilon_{ij} \partial_i \vec
\phi \times \partial_j \vec \phi \right) .
$$
Further, we choose the vacuum vector $\hat n = (0,0,1)$ such that the
potential $V=2(1-\phi^3)$ depends only on $\phi^3$. We will, in fact, allow
for general potentials $V=V(\phi^3)$ at the moment, because the Bogomolny
bound can be easily found for them. 
Using this, we rewrite the energy functional like
follows,
$$
E = \frac{1}{2} \int d^2 x \left( \frac{1}{4} (\epsilon_{ij}\vec
\phi \cdot (\partial_i \vec \phi \times \partial_j \vec \phi ))^2
+  \mu^2 V( \phi^3 ) \right)
$$
$$
= \frac{1}{2} \int d^2 x \left( \left( \frac{1}{2}
\epsilon_{ij}\vec \phi \cdot (\partial_i \vec \phi \times
\partial_j \vec \phi ) \pm  \, \mu \sqrt{V
 } \right)^2 \right.
$$
\begin{equation} \label{dIW-bogo}
\left. \mp   \, \mu \sqrt{V }
\, \, \, \epsilon_{ij}\vec \phi \cdot (\partial_i \vec \phi \times
\partial_j \vec \phi ) \right) .
\end{equation}
It remains to demonstrate that the last term above is, indeed, a Bogomolny
energy which is bounded by the topological charge.
For this purpose we use the complex field $u$ together with its modulus and
phase $u=fe^{i\Sigma}\equiv \sqrt{F}e^{i\Sigma}$ instead of the unit vector 
field $\vec \phi$. 
In these terms the topological charge reads 
\begin{equation} \label{top-charge}
Q[u] = \frac{-1}{2\pi i} \int d^2 x \frac{\epsilon_{ij} u_i \bar
u_j}{(1+u\bar u)^2} = \frac{1}{2\pi} \int d^2 x
\epsilon_{ij}\frac{ F_i \Sigma_j}{(1+F)^2}.
\end{equation}
For the energy we find from (\ref{dIW-bogo}) 
$$
E \ge \mp 4\pi  \mu \int d^2 x \left[\frac{1}{2\pi}  \sqrt{V(F)}
\epsilon_{ij}\frac{F_i \Sigma_j}{(1+F)^2} \right] \equiv 4\pi \mu C_1 |Q|
$$
where the sign has to be chosen accordingly (upper sign for $Q>0$). Further,
$C_1$ is a constant which depends on $V$, and the last equality still
has to be proven. If we replace $V$ by one, then the result is obvious,
because the integrand is just the integrand of the topological charge
(\ref{top-charge}). Equivalently, this expression is just the pullback of the
area form on the target space $S^2$, normalized to one. The base space
integral gives the result $Q$ because the base space sphere is covered $Q$
times while the target space sphere is covered once. But this last result
continues to hold with the factor $\sqrt{V(F)}$ present, up to a constant
$C_1$. Indeed, we just have to find a new target space coordinate $\tilde F$
instead of $F$ such that
\begin{equation} \label{FFtilde}
\frac{\sqrt{V(F)}dF}{(1+F)^2} = C_1 \frac{d\tilde F}{(1+\tilde F)^2}.
\end{equation}
The constant $C_1$, and a second coordinate $C_2$ which is provided by the
integration of Eq. (\ref{FFtilde}), are needed to impose the two conditions
\begin{equation} \label{FFbound}
\tilde F(F=0) =0 \, , \quad \tilde F(F=\infty) =\infty
\end{equation}
which must hold if $\tilde F$ is a good coordinate on the target space $S^2$.  
Obviously, $C_1$ depends on the potential. Specifically, for the ``old''
potential
$$
V=2(1-\phi^3) = 4\frac{F}{1+F}
$$
we find
$$
2\sqrt{\frac{F}{1+F}}\frac{dF}{(1+F)^2} = C_1 \frac{d\tilde F}{(1+\tilde F)^2}
$$
$$
\frac{4}{3}\left( \frac{F}{1+F} \right)^\frac{3}{2} = -\frac{C_1}{1+\tilde F}
+ C_2
$$
and, finally, from the boundary conditions (\ref{FFbound}), $C_1 = C_2
=(4/3)$.

The corresponding Bogomolny equations are 
$$
\frac{1}{2} \epsilon_{ij}\vec \phi \cdot (\partial_i \vec \phi
\times \partial_j \vec \phi ) \pm  \mu \sqrt{V(\phi^3)} =0
$$
or, in terms of $u=fe^{i\Sigma}$, $F=f^2$,
$$
2\epsilon_{ij}[\partial_i (1+F)^{-1} ]\partial_j \Sigma \pm \mu
\sqrt{V(F)} =0.
$$
For $F=F(r)$, $\Sigma = n\varphi$ this just gives the two possible square
roots of the first order equations found in Section 3, as may be checked
easily. 

%%%%%%%%%%%%%%%%%%%%%%%%%%%%%%%%%%%%%%%%%%%%%%%%%%%%%%%%%%%
\subsection{Bogomolny bounds in the full baby Skyrme model}
%%%%%%%%%%%%%%%%%%%%%%%%%%%%%%%%%%%%%%%%%%%%%%%%%%%%%%%%%%%
Here we want to discuss the fact that the above Bogomolny bound may be
used immediately to derive a tighter bound also for the full baby
Skyrme model \cite{deI-W}. 
The important point here is that the quadratic, O(3)
sigma model part of the full baby Skyrme model, which is absent in
the GP model, has its own, independent Bogomolny bound
$$
E_{O(3)} \ge 4\pi |Q|
$$
in terms of the same winding number. For the full baby Skyrme
model (\ref{bS}) we get, therefore, in the case of the ``old''
potential, the improved Bogomolny bound
$$
E_{bS} = E_{O(3)} + E_{GP} \ge 4\pi |Q|(1+\frac{4}{3}\mu ).
$$
In the full baby Skyrme model, however, solutions no longer
saturate this bound, because they would have to obey the field
equations both of the O(3) model and of the GP model in order to
do so.
\begin{figure}[h!]
\begin{center}
\includegraphics[angle=270,width=0.9 \textwidth]{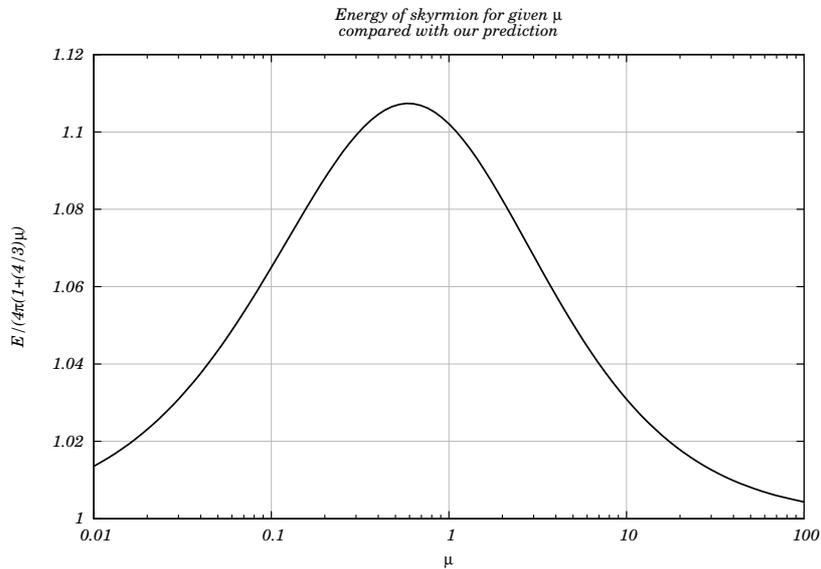}
\caption{Comparison of the energy and Bogomolny bound for $Q=1$
baby skyrmion}\label{rys1}
\end{center}
\end{figure}
\begin{figure}[h!]
\begin{center}
\includegraphics[angle=270,width=0.9 \textwidth]{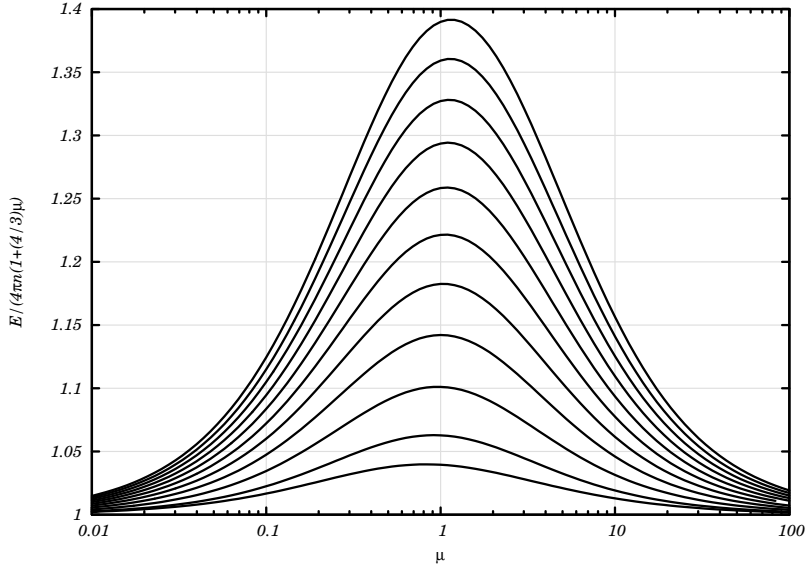}
\caption{Comparison of the energy and Bogomolny bound for
$Q=2..12$ baby skyrmions}\label{rys2}
\end{center}
\end{figure}
Nevertheless, we expect that the bound should be very tight for
very small and very large $\mu$. The reason for this is as
follows. A simple scaling argument shows that the potential and
the quartic pure Skyrme term always contribute the same amount of
energy to a solution. This implies that in the limit of very small
$\mu$, the energy approaches the pure quadratic O(3) sigma model
energy, whereas in the opposite limit of large $\mu$, the energy
approaches the energy of the GP model. But both the pure O(3)
model energy and the pure GP model energy saturate their
respective Bogomolny bounds. A simple numerical calculation of the
energies confirms this expectation. We make the rotationally
symmetric ansatz $u=f(r)e^{im\varphi}$ and determine $f(r)$ and
the corresponding energy by a simple shooting algorithm. The
deviation between energy and bound reaches its maximum of about
$10\%$ for values of $\mu \sim 1$, and diminishes both for small
and large $\mu$, as expected, see Fig. 1. The case $\mu = \sqrt{0.1}$ and
$Q=1$ corresponds to the numerical solution originally reported in
the first reference of \cite{old}, $E=1.564 \times 4\pi$, in which
case the new bound gives $1.422 \times 4\pi$ which is definitely
better than the old pure O(3) sigma model bound $4\pi$ ($10 \%$
off instead of more than $50 \%$ off).
\\
In Fig. 2 we show the corresponding 
results for higher topological charges $Q=2,...,12$.
Observe that although the symmetric ansatz does not necessarily
lead to the energy minima of the full baby Skyrme model, the bound
is again very tight in the large and small $\mu$ limits.

\vspace*{0.2cm}

The generalization to other potentials is obvious. The bound is
always of the type
$$
E_{bS} = E_{O(3)} + E_{GP} \ge 4\pi |Q|(1+C_1 \mu )
$$
where the numerical factor $C_1$ depends on the specific
potential but can be calculated exactly. We remark that the limit of large
$\mu$ has already been studied in \cite{Tchr 3}, although for a different
potential (the square of the ``old'' potential).
%%%%%%%%%%%%%%%%%%%%%%%%%%%%%%%%%%%%%%%%%%%%%%%%%%%%%%%%%%%%%%%%%
\section{Non-existence of non-topological $Q$-balls}
%%%%%%%%%%%%%%%%%%%%%%%%%%%%%%%%%%%%%%%%%%%%%%%%%%%%%%%%%%%%%%%%%
Here we show that there are no topologically trivial $Q$-balls in
the pure baby model with the old potential.
\\
We assume the symmetric ansatz for time-dependent solutions
\begin{equation}
u=e^{i  (\omega t + n\phi)} f(r). \label{ansatz t}
\end{equation}
Then, the equation of motion takes the form
$$
f\left[ -\frac{1}{r} \partial_r \left( \frac{rf'f}{(1+f^2)^2}
\left[ \frac{n^2}{r^2}-\omega^2\right] \right) +
\frac{1}{2}\right]=0.
$$
Thus, as always, a solution is given by the vacuum value $f=0$ or
by a solution of
$$
-\frac{1}{r} \partial_r \left( \frac{rf'f}{(1+f^2)^2} \left[
\frac{n^2}{r^2}-\omega^2\right] \right) + \frac{1}{2}=0.
$$
Now we introduce the $g$ function
$$
1-g=\frac{1}{1+f^2}
$$
and get the equation
\begin{equation}
-\frac{1}{r} \partial_r \left( rg' \left[
\frac{n^2}{r^2}-\omega^2\right] \right) + 1=0. \label{eq nontop}
\end{equation}
We are looking for non-topological solutions of the $Q$-ball type.
Hence, the relevant boundary conditions are: for a compacton
starting at $r=0$
$$
g(r=0)=0, \;\;\; g(r=R)=0, \;\;\; g'(r=R)=0
$$
or, for a shell-like compacton
$$
g(r=R_i)=0, \;\;\; g'(r=R_i)=0, \;\;\; i=1,2, \;\;\; R_1 < R_2.
$$
The general solution of the e.o.m. is
$$
g'(r)=\frac{1}{\frac{n^2}{r^2} -\omega^2} \left(\frac{r}{2} + A
\right)
$$
and
$$
g(r)=- \frac{r^2}{4\omega^2} - \frac{n^2}{4\omega^4} \ln
|n^2-\omega^2r^2|-\frac{A}{2\omega^2} \ln |n^2-\omega^2r^2|+B.
$$
First of all we may exclude shell-like compactons, as from
equation for $g'$ we get that
$$
g'(r=R)=0 \;\; \Rightarrow \;\; R^2=-2A.
$$
Hence, there is only one positive $R$. Notice, $A$ must be
negative. Further $g(0)=0$ gives
$$
B=\frac{n^2}{4\omega^4} \ln |n^2 |+\frac{A}{2\omega^2} \ln |n^2|
$$
and
$$
g(r)=- \frac{r^2}{4\omega^2} - \frac{n^2}{4\omega^4} \ln
\left|1-\frac{\omega^2r^2}{n^2}\right|-\frac{A}{2\omega^2} \ln
\left|1-\frac{\omega^2r^2}{n^2}\right| .
$$
The last constant $A$ can be found from $g(R)=0$
$$
0=- \frac{R^2}{4\omega^2} - \frac{n^2}{4\omega^4} \ln
\left|1-\frac{\omega^2R^2}{n^2}\right|-\frac{A}{2\omega^2} \ln
\left|1-\frac{\omega^2R^2}{n^2}\right| .
$$
But $R^2=-2A$, thus (after multiplication by $4\omega^2/n^2$
$$
0= \frac{2A\omega^2}{n^2} -  \ln
\left|1+\frac{2A\omega^2}{n^2}\right|-\frac{2A\omega^2}{n^2} \ln
\left|1+\frac{2A\omega^2}{n^2}\right|
$$
or
$$
0=p -  (1+p) \ln \left|1+p\right|,
$$
where $p=2A\omega^2/n^2$. This equation has two solutions $p=0$
and $p=p_0 < -1$. But in order to have a non-singular solution
$g$, the radial coordinate cannot be larger than $r<n/w$. So, the
size of the compacton must be also smaller than this critical
value $R<n/w$. As we know $R=\sqrt{-2A}$, where $2A=pn^2/w^2$ if
we use the definition of $p$. Then, $R=\sqrt{-p}n/w < n/w$. Thus
finally, $0<\sqrt{-p}<1$ i.e., $ p \in (-1,0)$, which is in 
contradiction with our previous observation that $p_0<-1$.

\vspace*{0.2cm}

The generalization to other potentials is unfortunately difficult. 
We are left with a quite complicated
second order nonlinear ODE for $g$. Fortunately, one may prove that for any
old baby potential (\ref{old s}) there are no non-spinning ($n=0$), 
non-topological
$Q$-balls. Instead of equation (\ref{eq nontop}), we get
$$
\frac{1}{r} \partial_r \left( rg' \right) + g^{s-1}=0,
$$
where the radial coordinate has been properly re-scaled. Assuming a series
expansion in the vicinity of the origin
$$ g(r)=Ar^a+...$$
we get
$$a=\frac{2}{2-s}, \;\;\;\; a^2+A^{s-2}=0. $$
The second formula leads to a complex or negative value of $A$, which is 
unacceptable as
$g$ must be positive. A similar effect occurs if we try an expansion at 
any finite $R$.
Then we have $$a(a-1)+A^{s-2}=0. $$
We can get an acceptable $A$ if $0<a<1$. But this gives $s<0$, 
which is excluded for obvious reasons.
Hence, no $n=0$, $Q$-balls are possible.

\vspace*{0.2cm}

The non-existence of non-topological $Q$-balls in the pure baby
Skyrme models is a rather new feature which has no analog in the full
baby Skyrme model. On the contrary, in the full theory one expects to find 
non-topological
time dependent solutions, compact or non-compact, depending on the
chosen form of the potential \cite{comp-bS}. Moreover, this result is also
unexpected, as topological $Q$-balls have been found by Gisiger and
Paranjape \cite{GP}, with properties similar to $Q$-baby skyrmions.
Apparently, the restricted model leads to non-trivial solutions which 
reflect properties of the full model only for field configurations with 
a non-trivial topological charge.

%%%%%%%%%%%%%%%%%%%%%%%%%%%%%%%%%%%%%%%%%%%%%%%%%%%%%%%%%%%%%%%%
\section{Conclusions}
%%%%%%%%%%%%%%%%%%%%%%%%%%%%%%%%%%%%%%%%%%%%%%%%%%%%%%%%%%%%%%%%
In this paper the pure baby Skyrme model has been analyzed. It was our main
aim to investigate whether solutions of this simplified model may tell us
something about the baby skyrmions, i.e., solitons in the full baby Skyrme
theory. Such a correspondence between solutions of the two models is by no 
means
obvious. Due to the huge amount of symmetries in the restricted model,
one could rather
expect a completely different behavior of solutions. 
We found, nevertheless, that
many properties of baby skyrmions are sufficiently well described by solutions
of the pure model. Moreover, as the restricted model is integrable
in the $(2+0)$ dimensional case (in $(2+1)$ dimensions the model
has the property of generalized integrability), 
we were able to perform all calculations analytically.
\\
The main findings are as follows:
\begin{enumerate}
\item[$\bullet$] Topological (charge) as well as geometrical
(nucleus/shell) features of baby skyr\-mions are captured already by
the soliton solutions of the pure model
\item[$\bullet$] Energies of baby skyrmions are reasonably approximated
by the energies of the solitons of the pure model. The approximation
becomes better and better while $\mu \rightarrow \infty$
\item[$\bullet$] There is a coincidence between the existence
of compact solitons in the pure baby model and the existence 
(for some values of the
parameters) of multi-baby-skyrmions in the full baby Skyrme model
\item[$\bullet$] There are no topologically trivial $Q$-ball configurations
\end{enumerate}
The first three results show that the static properties of baby skyrmions
may be qualitatively and quantitatively described by solutions of the
pure model. They indicate that the topological and geometrical properties
are governed by the Skyrme term and the potential, while the quadratic part of
the full baby Skyrme model only quantitatively modifies them.
Especially, the third result can be potentially important
as it gives a chance for an analytical treatment of the issue of the stability
of multi skyrmions. So the purely quartic
model, despite its high symmetry and probable integrability,
shares some properties with the full model, and may
probably be used to approximate the full model in some sense (for approximation
of solitons by compactons in the case of a scalar field theory see \cite{lis}).
\\
The fact that no topologically trivial $Q$-balls exist in the restricted model
leads to a different conclusion. $Q$-balls in the full baby Skyrme model are not
approximated by any solutions of the pure baby model but rather by solutions
of the $CP^1$ model, i.e., another simplified version of the baby Skyrme theory,
which consists of the quadratic part and potential. Of course, such a model
does not allow for static solutions and, therefore, is useless in the
context of the static baby skyrmions.
\\
The identification of simplified models relevant for the study of properties
of topological solitons (the pure baby Skyrme model) 
and non-topological solutions
(the $CP^1$ model with potential) is another result of the paper.

\section*{Acknowledgements}

C.A. and J.S.-G. thank MCyT (Spain), FEDER (FPA2005-01963) and
 Xunta de Galicia (grant INCITE09.296.035PR and
Conselleria de Educacion) for financial support. 
A.W. acknowledges support from the
Ministry of Science and Higher Education of Poland grant N N202
126735 (2008-2010). Further, A.W. thanks Ya. Shnir for discussion.

\end{document}